\documentclass[cits]{PoS}
\usepackage{amsmath,amssymb}
\usepackage{graphicx}

\newcommand{\Feff}{{\cal F}_\mathrm{eff}}
\newcommand{\Fint}{{\cal D}}
\newcommand{\chibar}{{\bar{\chi}}}
\newcommand{\psibar}{{\bar{\psi}}}
\newcommand{\eff}{\mathrm{eff}}
\newcommand{\comment}[1]{}
\title{
Another mean field treatment\\
in the strong coupling limit of lattice QCD
\footnote{Report No.: YITP-11-43}
}

\ShortTitle{Another mean field in SCL-LQCD}

\author{\speaker{Akira Ohnishi}\\
	Yukawa Institute for Theoretical Physics, 
        Kyoto University, Kyoto 606-8502, Japan\\
        E-mail: \email{ohnishi@yukawa.kyoto-u.ac.jp}}

\author{Kohtaroh Miura\\
        INFN Laboratori Nazionali di Frascati, 
	I-00044, Frascati (RM), Italy}

\author{Takashi Z. Nakano\\
	Department of Physics, Faculty of Science, Kyoto University, 
	Kyoto 606-8502, Japan\\
        Yukawa Institute for Theoretical Physics, 
	Kyoto University, Kyoto 606-8502, Japan}

\abstract{
We discuss the QCD phase diagram
in the strong coupling limit of lattice QCD
by using a new type of mean field coming
from the next-to-leading order of the large dimensional expansion.
The QCD phase diagram in the strong coupling limit
recently obtained by using the monomer-dimer-polymer (MDP) algorithm
has some differences in the phase boundary shape
from that in the mean field results.
As one of the origin to explain the difference,
we consider another type of auxiliary field,
which corresponds to the point-splitting mesonic composite.
Fermion determinant with this mean field
under the anti-periodic boundary condition
gives rise to a term which interpolates the effective potentials
in the previously proposed zero and finite temperature mean field treatments.
While the shift of the transition temperature at zero chemical potential
is in the desirable direction
and the phase boundary shape is improved,
we find that the effects are too large to be compatible
with the MDP simulation results.
}
\FullConference{The XXVIII International Symposium on Lattice Field Theory, Lattice2010\\
		June 14-19, 2010\\
		Villasimius, Italy}
\begin{document}
\section{Introduction}
The strong coupling lattice QCD (SC-LQCD) has been successful
from the beginning of the lattice QCD formulation.
In pure Yang-Mills theory,
the strong coupling limit (SCL) expression of the string tension
was proven to follow the area law,
and the Monte-Carlo simulation results on the string tension were
qualitatively explained in the strong coupling region
by including various plaquette configurations~\cite{Early}.
Recently SC-LQCD for pure Yang-Mills theory
is extended to finite temperatures~\cite{Langelage}.
SC-LQCD is also powerful in describing the QCD phase diagram.
The effective potential including fermion contributions
were obtained as a function of the chiral condensate in SCL,
and the spontaneous chiral symmetry breaking in vacuum
and its restoration at finite temperature and density
have been discussed based on the derived
effective potential~\cite{KS81,SCL-PhaseTransition,DHK1985,SCL-Std}.
Recently, SC-LQCD framework with fermions is extended 
to include the next-to-leading order
(NLO, ${\cal O}(1/g^2)$)~\cite{Faldt1986,NLO} 
and the next-to-next-to-leading order
(NNLO, ${\cal O}(1/g^4)$)~\cite{NNLO} 
contributions of the strong coupling expansion.
The mean field approach in SC-LQCD
overestimates the MC results of the transition temperature ($T_c$)
by around 10 \% in SCL,
while we overestimate $T_c$ by about 50-60 \% at $\beta =2N_c/g^2 \sim 4$.
When we include the Polyakov loop effects in SC-LQCD,
it is possible to roughly reproduce $T_c$
or the critical coupling ($\beta_c$) at a given temporal lattice size
($N_\tau = 1/T$) in the coupling region
$\beta=2N_c/g^2 \lesssim 4$~\cite{PSC-LQCD}.

Having these successes in mind, we expect that
the phase boundary predicted in the mean field treatment of SC-LQCD 
is promising in the strong coupling region.
Recently de Forcrand and Fromm~\cite{FF2010}
obtained the phase diagram in SCL-LQCD
in a Monte-Carlo simulation
based on the monomer-dimer-polymer (MDP) algorithm~\cite{MDP}.
In the MDP simulation, 
we first integrate out the link variables in SCL
(so-called zero $T$ treatment),
and the integral over quarks is simulated by the sum over loop configurations.
Since the gauge integral is carried out analytically,
the sign problem is weakened.
The shape of the phase boundary is somewhat different
from that in the mean field predictions;
$T_c$ for a given $\mu$ is much lower in the MDP results.
It is important to understand the origin of this deviation,
since the MDP simulation is only applicable to SCL ($1/g^2=0$)
and the mean field treatment of SC-LQCD is one of the few approaches
in which we can discuss cold dense matter directly based on 
non-perturbative QCD.

%
The deviation suggests that some of the approximations adopted
in the mean field treatment of SC-LQCD would not be good enough at finite $\mu$.
There are two types of approximations in the mean field approach.
One of them is 
the assumption that the auxiliary chiral field takes a constant value,
and the other is 
the truncation in the large dimensional ($1/d$) expansion~\cite{KlubergStern}.
In this proceedings,
we discuss the possibility to introduce another type of mean field
from the NLO term of the $1/d$ expansion
than the chiral condensate in SCL.
Specifically, we examine the role of the mean field of the type
$V_{\pm \nu, x}=\eta_{\nu,x}\bar{\chi}_x\chi_{x+\hat{\nu}}$
in the zero $T$ treatment.

\section{Strong coupling limit of lattice QCD with another mean field}
\label{sec:Feff}

In this section,
we first review the framework of the mean field treatment of SCL-LQCD
briefly, and compare the obtained phase diagram with that in the MDP simulation.
Next we introduce another kind of mean field.
Throughout this proceedings, we set the lattice spacing as $a=1$.

We consider the lattice QCD action with one species of unrooted staggered
fermion for color $\mathrm{SU}(N_c=3)$,
\begin{align}
\label{Eq:ZLQCD}
{\cal Z}_{\mathrm{LQCD}} 
=& \int \Fint[\chi,\chibar,U_\nu]~e^{-S_F-S_G}
\ ,\\
S_F
=&\frac {1}{2} \sum_x \sum_{\nu=0}^d 
  \left[ \eta_{\nu,x} \bar{\chi}_x U_{\nu,x} \chi_{x + \hat{\nu}} 
  - \eta_{\nu,x}^{-1} \bar{\chi}_{x + \hat{\nu}}
   U_{\nu,x}^{\dagger} \chi_x \right]
+m_0 \sum_x \bar{\chi}_x \chi_x 
\label{Eq:SCLaction}
,
\end{align}
where
$\chi (\bar{\chi})$, $m_0$, $U_{\nu,x}$,
$\eta_{\nu,x}=\exp(\delta_{\nu 0}\mu)(-1)^{x_0+\cdots +x_{j-1}}$
denote
the quark (antiquark) field,
the bare quark mass,
link variable,
and 
the staggered phase factor,
respectively.
The pure Yang-Mills action $S_G$ is proportional
to $1/g^2$, and disappears in SCL, $g \to \infty$.

\begin{figure}[thb]
\includegraphics[width=7.5cm]{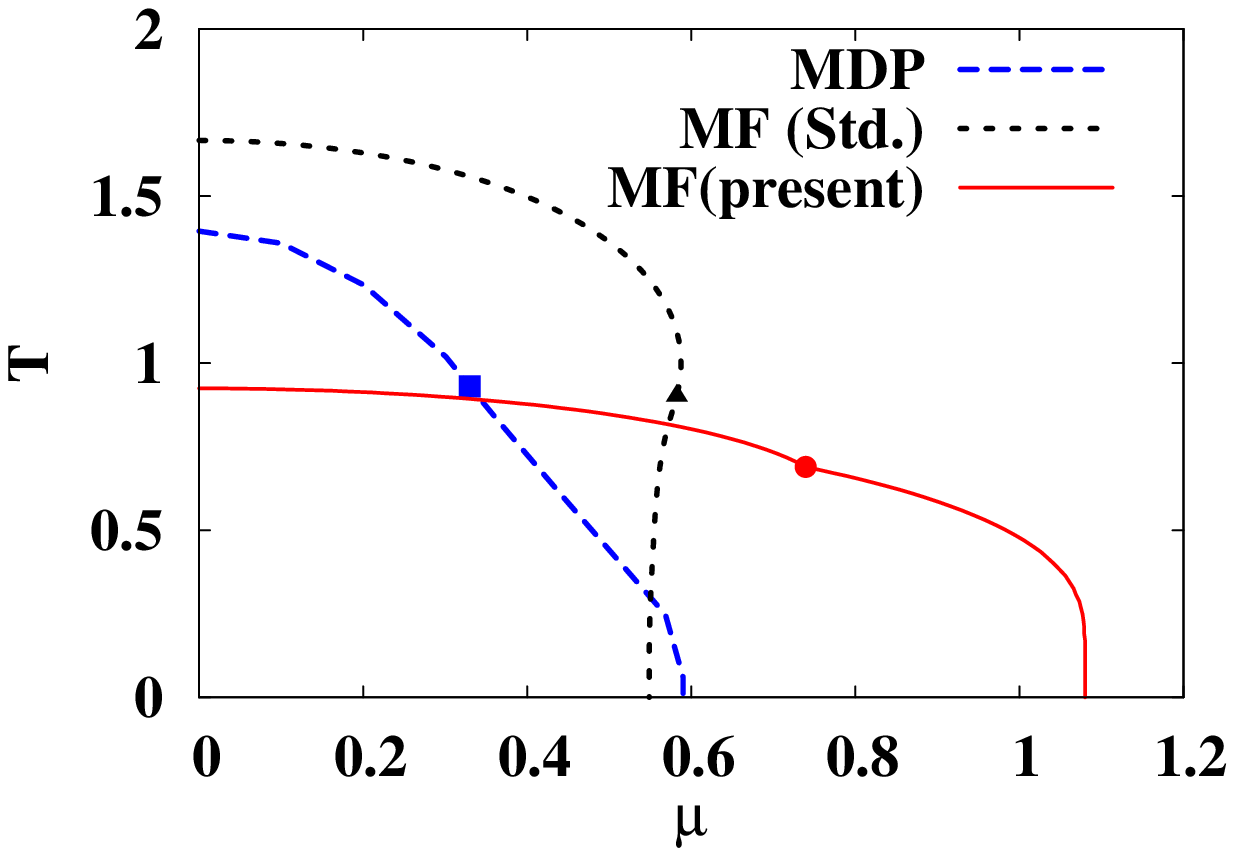}~%
\includegraphics[width=7.5cm]{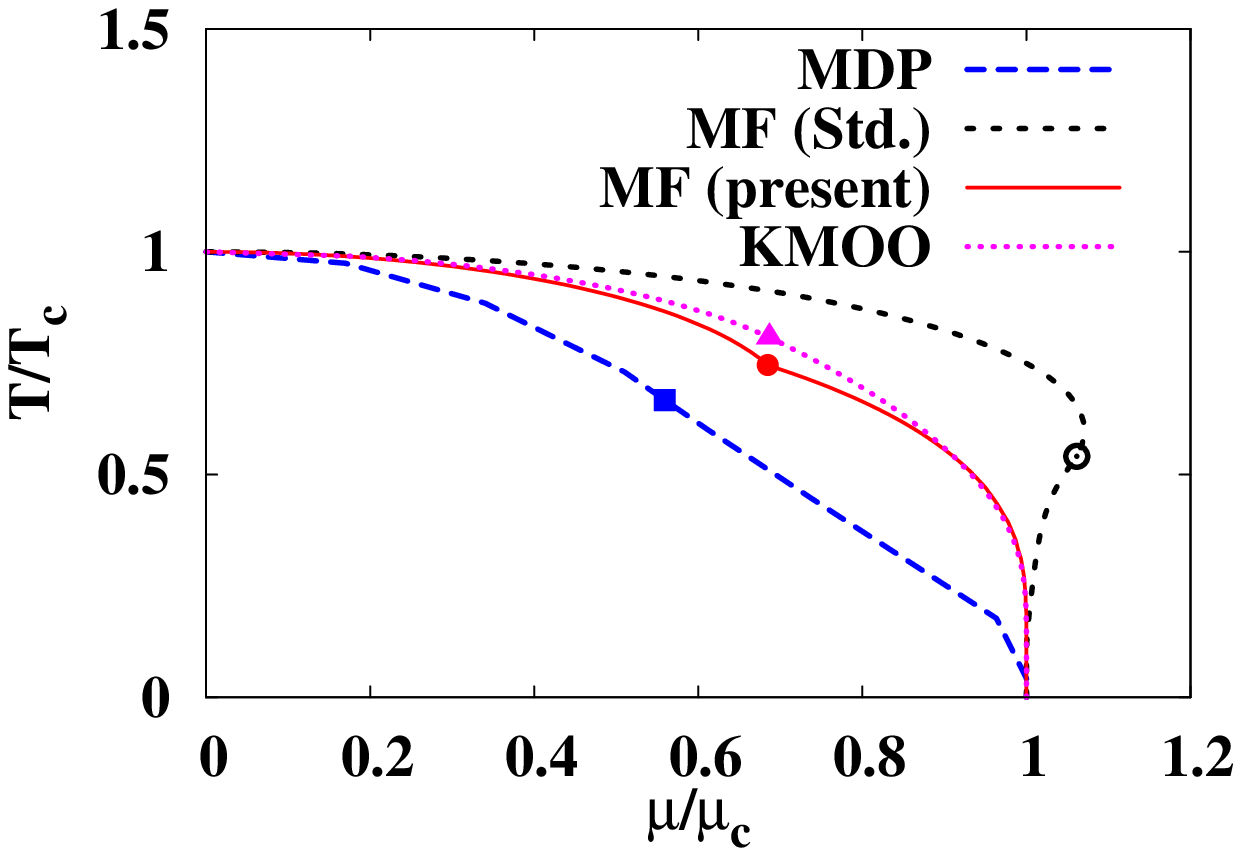}
\caption{Phase boundaries in 
the MDP simulation (dashed lines)~\cite{FF2010},
the standard mean field treatment (dotted lines)~\cite{SCL-Std},
and the present treatment (solid lines).
Left (right) panel shows the phase boundary in the lattice unit
(normalized by $T_c(\mu=0)$ and $\mu_c(T=0)$).
}
\label{Fig:pb}
\end{figure}

In the finite $T$ treatment of SCL-LQCD,
referred to as the standard mean field MF~(Std.) in the later discussion,
we first integrate out spatial link variables,
and obtain the effective action of quarks and the temporal links~\cite{SCL-Std},
\begin{align}
S^{\mathrm{(Std.)}}_\eff =& 
{1 \over 2} \sum_{x} \left[
e^\mu \chibar_x U_{0,x} \chi_{x+\hat{0}}
-e^{-\mu} \chibar_{x+\hat{0}} U^\dagger_{0,x} \chi_{x}
\right]
-{1 \over 4N_c} \sum_{\nu,x} M_x M_{x+\hat{\nu}}
+ m_0 \sum_x M_x
+ {\cal O}(1/\sqrt{d})
\ ,
\end{align}
where 
$d=3$ is the spatial dimension,
and $M_x= \chibar_x \chi_x$ represents a mesonic composite.
We next introduce an auxiliary field for the chiral condensate
to make the effective action bilinear in fermions,
and integrate fermions and temporal link variables,
for which the anti-periodic and periodic boundary conditions are imposed,
respectively.
The effective potential as a function of the chiral auxiliary field $\sigma$
is obtained in the mean field approximation as,
\begin{align}
\Feff^{\mathrm{(Std.)}}=
{d \over 4N_c} \sigma^2 - T \log \left[
{\sinh((N_c+1) E_q/T) \over \sinh(E_q/T)}
+ 2 \cosh (N_c\mu/T)
\right]
\label{Eq:FeffStd}
\ ,
\end{align}
where $E_q=\mathrm{arcsinh} (d\sigma/2N_c + m_0)$
is the one-dimensional quark excitation energy.
The phase diagram in MF~(Std.)
is shown by the dotted curve in Fig.~\ref{Fig:pb}.
Compared with the MDP results,
we overestimate the transition temperature at $\mu=0$ by 10-20 \%,
while we roughly reproduce the critical chemical potential at $T=0$.
In between,
the transition temperature stays high in the chemical potential region
$\mu < 0.5$ in MF~(Std.),
and the phase boundary shape is significantly different from that in MDP.

There are two types of approximations adopted in MF~(Std.).
One of them is the assumption that the introduced auxiliary field
$\sigma$ is constant,
and the other one is the truncation in the $1/d$ expansion.
In the $1/d$ expansion~\cite{KlubergStern},
the $M_x M_{x+\hat{\nu}}$ term containing four quarks and sum over dimensions
is assumed to be finite at large $d$.
Then the quark field scales as $d^{-1/4}$,
and the NLO terms containing six quarks
are proportional to $1/\sqrt{d}$.
Baryonic composite action belongs to NLO,
and it is natural to expect
that the baryonic action would affect the phase boundary at finite $\mu$.
Baryonic composite effects have been discussed at finite $T$
in Refs.~\cite{KMOO2007}, but the adopted bosonization method~\cite{Azcoiti}
is not fully compatible with the chiral symmetry.
Further studies are necessary to discuss the chiral phase transition seriously.

By comparison with MF~(Std.),
both spatial and temporal link variables are integrated out first
in the zero $T$ treatment.
The effective action including the NLO terms in the $1/d$ expansion
is given as,
\begin{align}
S_\eff =& -{1 \over 4N_c} \sum_{\nu,x} M_x M_{x+\hat{\nu}}
+ m_0 \sum_x M_x
+{1\over 8}\sum_{\nu,x}
\left[
  \eta_{\nu,x}^3 \bar{B}_x B_{x+\hat{\nu}}
- \eta_{\nu,x}^{-3} \bar{B}_{x+\hat{\nu}} B_x
\right]
+{\cal O}(1/(d+1))
\label{Eq:ZeroTaction}
\ ,
\end{align}
where 
$B_x= \varepsilon_{abc} \chi^a_x \chi^b_x \chi^c_x/6$
is the baryonic composite.
The third term in Eq.~(\ref{Eq:ZeroTaction}) 
is the NLO term, ${\cal O}(1/\sqrt{d})$, in the $1/d$ expansion;
it contains six fermions ($\propto d^{-3/2}$)
and sum over space-time dimensions.
This effective action with higher order terms in the large dimensional 
expansion is used in the MDP simulation~\cite{MDP}.
In Ref.~\cite{DHK1985},
the phase transition at finite $\mu$ and zero $T$ was investigated
by introducing the auxiliary baryon field $b \sim B$
in the mean field approximation for the chiral condensate,
but the phase transition at finite $T$ is not well described.
This shortcoming could come
from the lack of the anti-periodic boundary condition,
which is decisive in the chiral transition at finite $T$.

In order to impose the anti-periodic boundary condition for quarks
in the effective action Eq.~(\ref{Eq:ZeroTaction}),
we here consider another type of auxiliary field,
which corresponds to the point-splitting mesonic composite,
\begin{align}
V_{+\nu,x} = \eta_{\nu,x}\chibar_x \chi_{x+\hat{\nu}}\ , 
\quad
V_{-\nu,x} = \eta_{\nu,x}^{-1}\chibar_{x+\hat{\nu}} \chi_{x}\ .
\end{align}
We note that the baryonic term in Eq.~(\ref{Eq:ZeroTaction})
can be rewritten by using the anti-commuting property of the Grassmann 
variables as
\begin{align}
\eta_{\nu,x}^3 \bar{B}_x B_{x+\hat{\nu}}
=(V_{+\nu,x})^3/6
\ ,\quad
\eta_{\nu,x}^{-3} \bar{B}_{x+\hat{\nu}} B_x
=(V_{-\nu,x})^3/6
\ .
\end{align}
By applying the extended Hubbard-Stratonovich transformation~\cite{NLO},
\begin{align}
\exp(\mp\alpha V^3) \simeq& \exp\left[
	-\alpha(\psibar^{(1)}\psi^{(1)}-V^2 \psi^{(1)}\pm\psibar^{(1)}V)
	\right]
\ ,\\
\exp(\alpha\psi^{(1)}V^2) \simeq& \exp\left[
	-\alpha(\psibar^{(2)}\psi^{(2)}-V \psi^{(2)}-\psibar^{(2)}V\psi^{(1)})
	\right]
\ .
\end{align}
we obtain the effective action for quarks
and auxiliary fields,
\begin{align}
S_\mathrm{eff}
=&N_\tau L^d \Feff^{(X)}(\sigma, \psi^{(k)}_{\pm\nu}, \psibar^{(k)}_{\pm\nu})
+\frac12 \sum_{\nu,x}\left[
	Z_{+\nu}V_{+\nu,x}-Z_{-\nu} V_{-\nu,x}
	\right]
+m_q \sum_x M_x
\ ,\\
Z_{\pm\nu}=&2\alpha\left(\psibar^{(1)}_{\pm\nu} \mp \psi^{(2)}_{\pm\nu}
	\mp \psibar^{(2)}_{\pm\nu}\psi^{(1)}_{\pm\nu}
	\right)
\ .
\end{align}
Under the assumption that auxiliary fields take constant values,
we can evaluate the Matsubara product
with the anti-periodic boundary condition for quarks.
Equilibrium condition for $\psi^{(k)}_{\pm\nu}$ and $\psibar^{(k)}_{\pm\nu}$
is used to reduce the number of independent variables.
For example, equilibrium values of 
$\psi^{(k)}_{+\nu}$ and $\psibar^{(k)}_{+\nu}$ are related as,
$\psibar^{(2)}_{+\nu}=-\psi^{(1)}_{+\nu}\equiv\varphi_{+\nu}$
and
$\psibar^{(1)}_{+\nu}=-\psi^{(2)}_{+\nu}=\varphi^2_{+\nu}$.
We also utilize the rotational and reflection symmetry,
$\varphi_{+j}=\varphi_{-j}\equiv\varphi_s (j=1,2,3)$.
The effective potential is found to be as follows.
\begin{align}
\Feff=&
\frac{d+1}{4N_c} 
\sigma^2 + 2\alpha (\varphi_+^3 + \varphi_-^3)
	+ 4\alpha d \varphi_s^3 + V_q
\quad (\alpha=1/48)
\ ,\\
V_q=&
-N_c T \frac{1}{L^d} \sum_k
	\left[
	\frac{E_k}{T}
	+ \log\left( 1 + e^{-(E_k - \tilde{\mu})/T}\right)
	+ \log\left( 1 + e^{-(E_k + \tilde{\mu})/T}\right)
	\right]
- N_c \log Z_\chi
\label{Eq:Vq}
\ ,\\
Z_\pm =& 6\alpha\varphi_\pm^2
\ ,\quad
Z_\chi = \sqrt{Z_+Z_-}
\ ,\quad
\tilde{\mu}=\mu + \log(Z_+/Z_-)
\ ,\\
E_k=&\mathrm{arcsinh} (\varepsilon_k/Z_\chi)
\ ,\quad
\varepsilon_k=\sqrt{m_q^2+Z_s^2 \sin^2 \bold{k}}
\ ,\quad
m_q = \frac{d+1}{2N_c}\sigma + m_0
\ ,\quad
Z_s = 6\alpha\varphi_s^2
\label{Eq:EffPotZ}
\ .
\end{align}
This effective potential has interesting features;
quarks couple with auxiliary fields via the constituent quark mass $m_q$
and the wave function renormalization factor $Z_{\pm,s}$,
and it contains the momentum integral
with a lattice type replacement, $k \to \sin k$.

It should be noted that
the composite $V_{\pm\nu}$ is not gauge invariant,
and the mean field introduced here should be regarded
as the one in a fixed gauge.
We expect that this feature may not be serious,
since we are discussing the dynamics in the link integrated effective action
Eq.~(\ref{Eq:ZeroTaction}).

\section{Effective potential surface and phase boundary}\label{sec:res}

We shall now examine the effective potential Eq.~(\ref{Eq:EffPotZ}).
For simplicity, we ignore the effects of $\varphi_s$ which connects
spatially separated quarks,
and we discuss the results only in the chiral limit, $m_0=0$.
We first consider the $\Feff$ in vacuum, $(T,\mu)=(0,0)$.
At $\mu=0$, the temporal forward and backward auxiliary fields
take the same equilibrium value, $\varphi_+=\varphi_-$.
When $\varphi_\pm$ are zero, the vacuum effective potential
becomes the logarithmic type,
$\Feff \to b_\sigma \sigma^2/2 - N_c \log m_q$,
which is the leading order effective potential of the $1/d$ expansion
in the zero $T$ treatment~\cite{KS81}.
For finite $\varphi_\pm$,
the effective potential becomes the arcsinh type,
which is typical in the finite $T$ treatment.
Thus the present effective potential may be regarded as an interpolating
one of the zero and finite $T$ effective potentials.

\begin{figure}[ht]
\centerline{
\includegraphics[width=15cm]{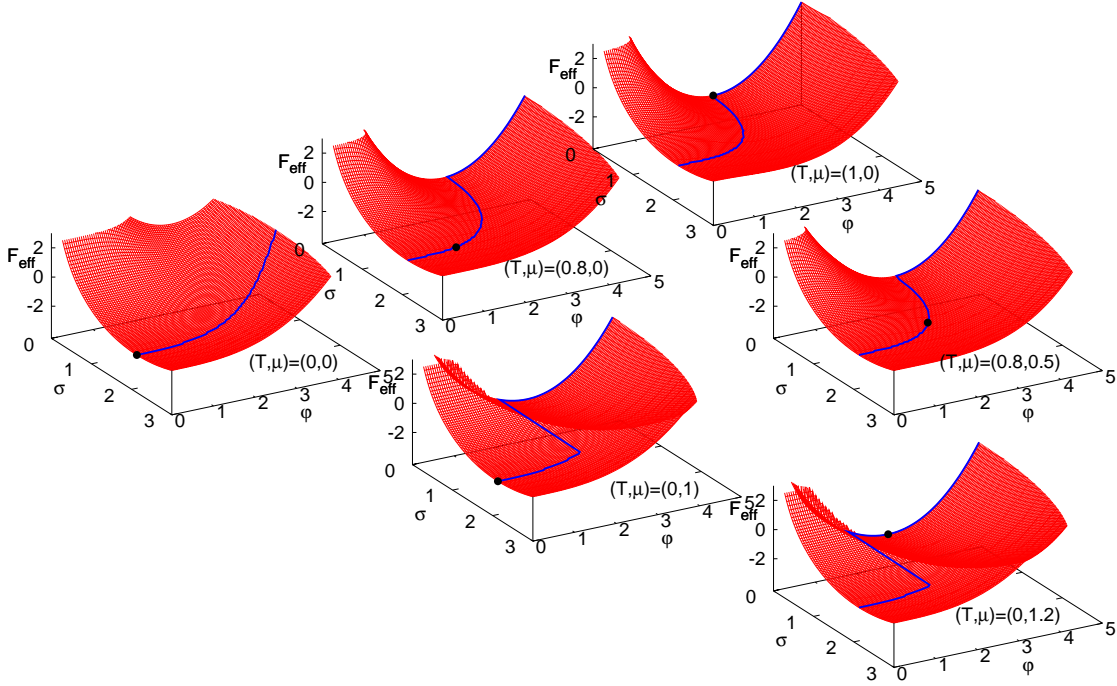}~%
}
\caption{Effective potential surface in the $(\sigma, \varphi=\sqrt{\varphi_+\varphi_-})$ plane.
The solid line connects the points which give the lowest $\Feff$ for a given $\varphi$ value,
and the filled circles show the equilibrium.
}
\label{Fig:surf}
\end{figure}

In Fig.~\ref{Fig:surf}, we show the effective potential $\Feff$
as a function of $\sigma$ and $\varphi\equiv\sqrt{\varphi_+\varphi_-}$
at several $(T,\mu)$.
In vacuum, equilibrium is realized at finite $\sigma$ and $\varphi_\pm=0$,
while $\varphi_\pm$ ($\sigma$) grows (decreases) as $T$ increases.
The phase transition at finite $T$ and zero $\mu$ is the second order.
The transition temperature at $\mu=0$ ($T_c=0.92$) is smaller
than those in MDP ($T_c\simeq 1.4$) and MF~(Std.) ($T_c=5/3$).
The decrease of $T_c$ may be understood as the contribution from single quarks.
In MF~(Std),
coherently moving three quarks contribute to the effective potential as a baryon
as seen in the Boltzmann factor of $\exp(-N_c \mu/T)$ in Eq.~(\ref{Eq:FeffStd}).
By comparison, the mean field $\varphi_\pm$ allows a single quark
excitation as found in the Boltzmann factor $\exp(-\mu/T)$ in Eq.~(\ref{Eq:Vq}).

At finite $\mu$ and $T=0$,
the energy surface is separated by the ridge at $E_q=\tilde{\mu}$,
and the vacuum configuration (finite $\sigma$ and zero $\varphi$)
jumps to the high density configuration (zero $\sigma$ and finite $\varphi$).
The transition chemical potential at $T=0$ ($\mu_c = 1.08$) is larger
than those in MDP ($\mu_c\simeq 0.59$) and MF~(Std.) ($\mu_c=0.55$).

In Fig.~\ref{Fig:pb}, we compare the phase boundary
in the present treatment including $\varphi_\pm$
with those in the MDP simulation and MF~(Std.).
The new mean fields $\varphi_\pm$ shift $T_c$ and $\mu_c$
from those in MF~(Std.) in right directions,
and when we normalize $T$ and $\mu$ by $T_c(\mu=0)$ and $\mu_c(T=0)$,
the phase boundary shape is significantly improved
and becomes is similar to that
in the finite $T$ treatment with baryonic composite effects~\cite{KMOO2007}.
However, their effects are too much in the shifts of $T_c$ and $\mu_c$.

\section{Summary}\label{sec:sum}

In this proceedings, we have discussed how we can understand
the phase boundary in the Monomer-Dimer-Polymer (MDP) simulation~\cite{FF2010}
in the strong coupling limit (SCL) of lattice QCD for color SU(3)
with unrooted staggered fermion
in an analytical method based on the mean field treatment.
Since the MDP simulation is available at present only in SCL,
it is important to understand it in a method which is applicable
to finite coupling cases.
Here we have examined a new type of mean field $\varphi_\pm$,
which connects the fermion in a different temporal variable
in the zero $T$ treatment,
where both spatial and temporal link integrals are carried out first.
This enables us to impose the anti-periodic boundary condition of quarks,
and may be important to describe the phase transition at finite $T$.
Actually we can describe the finite $T$ phase transition
in the zero $T$ treatment.
The new mean field shifts the transition temperature at $\mu=0$
and the transition chemical potential at $T=0$ in the right directions,
but we find that the effects are too much to be compatible
with the MDP results.
It is still a challenge to understand the phase diagram
in the MDP simulation.
It would be necessary to examine
the combination with the baryonic effects
proposed so far~\cite{DHK1985,KMOO2007}
and the fluctuation of the chiral condensate.

\section*{Acknowledgments}
We would like to thank 
Alejandro Vaquero and Professor Vicente Azcoiti
for useful discussions.
This work was supported in part
by Grants-in-Aid for Scientific Research from JSPS (Nos. 22-3314),
the Yukawa International Program for Quark-hadron Sciences (YIPQS),
and by Grants-in-Aid for the global COE program
``The Next Generation of Physics, Spun from Universality and Emergence''
from MEXT.

\end{document}